\documentclass[superscriptaddress,onecolumn]{revtex4}
\usepackage{amssymb}
\usepackage{amsfonts}
\usepackage{amsmath}
\usepackage{graphicx}
\usepackage{dcolumn}
\usepackage{bm}

\begin{document}


\title{Cosmological frames for theories with absolute parallelism}

\author{Rafael Ferraro}
\email{ferraro@iafe.uba.ar}
\thanks{Member of Carrera del Investigador Cient\'{\i}fico (CONICET,
Argentina)} \affiliation{Instituto de Astronom\'\i a y F\'\i sica
del Espacio, Casilla de Correo 67, Sucursal 28, 1428 Buenos Aires,
Argentina} \affiliation{Departamento de F\'\i sica, Facultad de
Ciencias Exactas y Naturales, Universidad de Buenos Aires, Ciudad
Universitaria, Pabell\'on I, 1428 Buenos Aires, Argentina}
\author{Franco Fiorini}
\email{franco@iafe.uba.ar} \affiliation{Instituto de Astronom\'\i
a y F\'\i sica del Espacio, Casilla de Correo 67, Sucursal 28,
1428 Buenos Aires, Argentina} 


\begin{abstract}
The vierbein (tetrad) fields for closed and open
Friedmann-Robertson-Walker cosmologies are hard to work out in most
of the theories featuring absolute parallelism. The difficulty is
traced in the fact that these theories are not invariant under local
Lorentz transformations of the vierbein. We illustrate this issue in
the framework of $f(T)$ theories and Born-Infeld determinantal
gravity. In particular, we show that the early Universe as described
by the Born-Infeld scheme is singularity free and naturally
inflationary as a consequence of the very nature of Born-Infeld
gravitational action. 
\end{abstract}

\maketitle


\section{Introduction}

In the last decade a wide variety of modified theories of gravity
has been studied with the aim of solving or smoothing some puzzling
features of conventional gravity and cosmology. The reasons for
considering such a modified gravitational schemes rely mainly on two
damages that are intrinsic features of general relativity (GR). On
one hand, it is widely accepted by high energy physicists that the
description of the gravitational field provided by general
relativity must be doomed at scales of the order of the Planck
length, where the spacetime structure itself must be represented in
terms of a quantum regime. On the other hand, and in the opposite
extreme of the physical phenomena, GR also faces an intriguing
dilemma in connection with the late cosmic speed up stage of the
Universe.

One of the newest extended theories of gravity is the so called
$f(T)$ gravity, which is a theory formulated in a spacetime
possessing absolute parallelism \cite{Hehl1,Hehl2}. The first
investigations within this framework can be traced back to
Refs.~\cite{Nos}--\cite{Nos3}, where an ultraviolet deformation of
Einstein gravity was considered with the aim of smoothing
singularities, namely the initial singularity of
Friedmann-Robertson-Walker (FRW) cosmological models. After that,
the attention was focused in low energy deformations of GR which
serve to tackle those aspects of the cosmological evolution
connected with the late speed up stage of the universe in
spatially flat models \cite{fdet1}--\cite{fdet10}. Quite more
recently, some fundamental aspects of $f(T)$ theories, like the
presence of extra degrees of freedom and the lack of Lorentz
invariance, had been addressed in
Refs.~\cite{fdetB1}--\cite{Nos5}. In spite of their success in
smoothing the Big-Bang singularity \cite{Nos}, the ultraviolet
deformations in the context of the $f(T)$ theories fail to smooth
singularities of GR vacuum solutions like black holes \cite{Nos2}.
For this reason, we introduced in Ref. \cite{Nos4} a more general
theory in which the singularities of Einstein theory receive a
more appropriate treatment, even those that emerge from vacuum
solutions of Einstein equations. In this framework the
gravitational action is chosen, in close analogy with the
Born-Infeld (BI) procedure used in electrodynamics, as the
determinant of a particular combination of the torsion tensor that
assures the correct low energy limit given by GR. Both the BI
theory and the $f(T)$'s mentioned above are based on actions built
with just first derivatives of the dynamical field -the vierbein
$e^a$-, so assuring second order motion equations.

Most of the gravitational theories with absolute parallelism lack
the invariance under local Lorentz transformations of the vierbein.
This means that they should be regarded as theories describing the
dynamics of the vierbein field rather than the metric field (which
is invariant under local Lorentz transformations of the vierbein).
As a consequence, the symmetry of the metric tensor is not enough to
suggest the form of the vierbein, which could be elusive in many
physically meaningful solutions. In this article we will examine
this issue by working out the proper vierbein field for closed and
open FRW cosmologies.  For this purpose, in Section 2 we introduce
the teleparallel equivalent of general relativity, that constitutes
the geometrical pillar on which the $f(T)$ and BI theories of
gravity, also summarized there, are built. In Section 3 we explain
the lack of invariance of these theories under local Lorentz
transformation of the vierbein field in a cosmological context,
discuss about the meaning of this feature, and work out the proper
vierbein field for closed and open FRW cosmologies. Finally, in
Section 4 we display the conclusions.

\section{Gravitational theories with absolute parallelism: two examples}

In this Section we will focus the attention on two gravitational
schemes based on absolute parallelism. On one hand, we will discuss
a family of modified teleparallel theories constructed as a
deformation of the Einstein-Hilbert Lagrangian in Weitzenb\"{o}ck
spacetime --the so called $f(T)$ gravities--. On the other hand, we
will consider a quite different ultraviolet modification of GR with
determinantal structure: Born-Infeld gravity. We will start by
introducing the geometrical concepts that serve as the building
blocks of both schemes.

The basic idea is that gravity can be described by providing the
spacetime with a torsion $T^{a}=de^{a}$, $a=0,...,3$, where
$\{e^{a}\}$ is a vierbein (a basis of the cotangent space) in a
$4$-dimensional manifold. The vierbein $\{e^{a}\}$ is the co-frame
of an associated basis $\{e_{a}\}$ in the tangent space. If
$e_{\mu }^{a}$ and $e_{a}^{\mu }$ are respectively the components
of the 1-forms $e^{a}$ and the vectors $e_{a}$\ in a given
coordinate basis, then the relation between frame and co-frame is
expressed as
\begin{equation}
e_{\mu }^{a}\ e_{b}^{\mu }=\delta _{b}^{a}\ .  \label{coframe}
\end{equation}
Contracting with $e_{a}^{\nu }$ one also gets
\begin{equation}
e_{a}^{\nu }\ e_{\mu }^{a}=\delta _{\mu }^{\nu }\ .
\label{coframe2}
\end{equation}
The components $T_{\ \ \mu \nu }^{\lambda }$\ of the torsion
tensor in the coordinate basis is related to the 2-forms $T^{a}$
through the equation
\begin{equation}
T_{\ \ \mu \nu }^{\lambda }\equiv e_{a}^{\lambda }\ T_{\ \ \mu \nu
}^{a}=e_{a}^{\lambda }\,(\partial _{\nu }e_{\mu }^{a}-\partial
_{\mu }e_{\nu }^{a})\ .  \label{torsion}
\end{equation}
This means that the spacetime is endowed with a connection
\begin{equation}
{\Gamma }_{\mu \nu }^{\lambda }=e_{a}^{\lambda }\,\partial _{\nu
}e_{\mu }^{a}+\text{terms symmetric in}\,\, \mu \nu,  \label{Wei}
\end{equation}
(since $T_{\ \ \mu \nu }^{\lambda }\equiv {\Gamma _{\nu \mu
}^{\lambda }}-{\Gamma _{\mu \nu }^{\lambda }}$). The first term in
Eq.~(\ref{Wei}) is the Weitzenb\"{o}ck connection. The metric is
introduced as a subsidiary field given by
\begin{equation}
g_{\mu \nu }(x)=\eta _{ab}\ e_{\mu }^{a}(x)\ e_{\nu }^{b}(x)\ ,
\label{metric}
\end{equation}
where $\eta _{ab}=diag(1,-1,-1,-1)$. Eq.~(\ref{metric}) can be
inverted with the help of Eq.~(\ref{coframe}) to obtain
\begin{equation}
\eta _{ab}=g_{\mu \nu }(x)\ e_{a}^{\mu }(x)\ e_{b}^{\nu }(x)\ ,
\label{orto}
\end{equation}
which means that the vierbein is orthonormal.

Teleparallelism uses the Weitzenb\"{o}ck spacetime, where the
connection is chosen as
\begin{equation}
{\Gamma }_{\mu \nu }^{\lambda }=e_{a}^{\lambda }\,\partial _{\nu
}e_{\mu }^{a}\ .  \label{Wei1}
\end{equation}
As a consequence of the choice of the Weitzenb\"{o}ck connection
(\ref{Wei1}), the Riemann tensor is identically null. So the
spacetime is flat: the gravitational degrees of freedom are
completely encoded in the torsion $T^{a}=de^{a}$.

In terms of parallelism, the choice of the Weitzenb\"{o}ck
connection has a simple meaning. In fact, the covariant derivative
of a vector yields
\begin{equation}
\nabla _{\nu }V^{\lambda }=\partial _{\nu }V^{\lambda }+{\Gamma
}_{\mu \nu }^{\lambda }V^{\mu } = e_{a}^{\lambda }\ \partial _{\nu
}(e_{\mu }^{a}V^{\mu })\equiv e_{a}^{\lambda }\ \partial _{\nu
}V^a\, .\label{transport}
\end{equation}
In particular, Eq.~(\ref{orto}) implies that $\nabla _{\nu
}e_{b}^{\lambda }=0$; so, the Weitzenb\"{o}ck connection is metric
compatible. In general, Eq.~(\ref{transport}) means that a given
vector is parallel transported along a curve if its projections on
the co-frame remain constant. So, the vierbein parallelizes the
spacetime. Of course, this nice criterion of parallelism would be
destroyed if local Lorentz transformations of the co-frame were
allowed in the theory.

Teleparallelism is a dynamical theory for the vierbein, which is
built from the torsion. According to Eq.~(\ref{metric}), a set of
dynamical equations for the vierbein also implies a dynamics for
the metric. This dynamics coincides with Einstein's dynamics for
the metric when the teleparallel Lagrangian density is chosen as
\cite{Haya,Maluf}
\begin{equation}
\mathcal{L}_{\mathbf{T}}[e^{a}]=\frac{1}{16\pi
G}\;e\;(T-2\Lambda)\;, \label{lagrangianT}
\end{equation}
where $e\equiv |e_{\mu }^{a}|=\sqrt{|g_{\mu \nu }|}$, and $|\,\,|$
stands for the absolute value of the determinant. In Eq.
(\ref{lagrangianT}) we have defined the \emph{Weitzenb\"{o}ck
invariant} as
\begin{equation}
T%
=S_{\ \mu \nu }^{\rho }T_{\rho }^{\ \mu \nu }. \label{Weitinvar}
\end{equation}
The tensor $S_{\ \mu \nu }^{\rho }$ appearing in the last equation
is defined according to
\begin{equation}
S_{\ \mu \nu }^{\rho }=\frac{1}{4}\,(T_{\ \mu \nu }^{\rho }-T_{\mu
\nu }^{\ \ \ \rho }+T_{\nu \mu }^{\ \ \ \rho })+\frac{1}{2}\
\delta _{\mu }^{\rho }\ T_{\sigma \nu }^{\ \ \ \sigma
}-\frac{1}{2}\ \delta _{\nu }^{\rho }\ T_{\sigma \mu }^{\ \
\,\sigma }\;.  \label{tensor}
\end{equation}
In fact, the Lagrangian (\ref{lagrangianT}) just differs from the
Einstein-Hilbert Lagrangian with cosmological constant
$\mathcal{L}_{\mathbf{GR}}=-(16\pi G)^{-1}\ e\ (R+2\Lambda)$\ in a
total derivative
\begin{equation}
-e\ R[e^{a}]=e\ T-2\;\partial _{\nu }(e\;T_{\sigma }^{\ \ \sigma
\nu }\,)\;,  \label{divergence}
\end{equation}
where $R$ is the scalar curvature for the Levi-Civita connection.
When GR dresses this costume, the gravitational degrees of freedom
are gathered in the torsion instead of the Levi-Civita curvature.
It is a very curious and fortunate fact that both pictures enable
to construct a gravitational action with the same physical
content. However, it is remarkable that the Lagrangian
(\ref{lagrangianT}) involves just first derivatives of its
dynamical field, the vierbein. In some sense, the teleparallel
Lagrangian picks up the essential dynamical content of Einstein
theory without the annoying second order derivatives appearing in
the last term of Eq.~(\ref{divergence}). Such Lagrangian is a
better starting point for considering modified gravity theories,
since any deformation of its dependence on $T$ will always lead to
second order dynamical equations. On the contrary, the so called
$f(R)$ theories lead to fourth order equations.

Analogously to the $f(R)$ scheme, a $f(T)$ theory replaces the
Weitzenb\"{o}ck invariant $T$ in Eq.~(\ref{lagrangianT}) with a
general function $f(T)$. So, the dynamics is described by the
action
\begin{equation}
\mathcal{I}=\frac{1}{16\pi G}\int d^{4}x\;e\;f(T)+\int
d^{4}x\;\mathcal{L}_{\mathcal{M}}\;.  \label{action}
\end{equation}%
where $\mathcal{L}_{\mathcal{M}}$ is the matter Lagrangian
density. Undoubtedly, the whole family of actions gathered in
(\ref{action}) constitutes a vast territory worth to be explored,
specially when one is aware that the dynamical equations arising
by varying the action (\ref{action}) with respect to the vierbein
components $e_{\mu }^{a}(x)$ are of second order. This distinctive
feature makes Weitzenb\"{o}ck spacetime a privileged geometric
structure to formulate modified theories of gravitation. In fact,
the dynamical equations for the vierbein tell how the matter
distribution organizes the orientation of the basis $e^{a}$ at
each point, in such a way that the field lines of $e^{a}(x)$
realize the parallelization of the manifold. After this vierbein
field is obtained, one uses the assumption of orthonormality to
get the metric (\ref{metric}).

 \bigskip

Let us deal now with Born-Infeld gravity. This high energy
modification of Einstein gravity is based on the Born-Infeld
procedure used first in the context of electrodynamics, and is
governed by the $n$-dimensional action in Weitzenb\"{o}ck
spacetime given by \cite{Nos4}
\begin{equation}
\mathcal{I}_{\mathbf{BIG}}=\frac{\lambda /(A\!+\!B)}{16\pi G}\!\int \!d^{n}x%
\left[ \sqrt{|g_{\mu \nu }+2\lambda ^{-1}\mathcal{F}_{\mu \nu
}|}-\alpha \sqrt{|g_{\mu \nu }|}\right],   \label{acciondet}
\end{equation}%
where $\mathcal{F}_{\mu \nu }$ is quadratic in the Weitzenb\"{o}ck
torsion, and reads
\begin{equation}\label{tensorF}
\mathcal{F}_{\mu \nu }=A\,S_{\mu \lambda \rho }T_{\nu
}^{\,\,\,\lambda \rho }+B\,S_{\lambda \mu \rho }T_{\,\,\,\,\,\nu
}^{\lambda \,\,\,\,\rho },
\end{equation}
being $A$ and $B$ non-dimensional constants. Such a
combination ensures the correct GR limit since both terms in $\mathcal{F}%
_{\mu \nu }$ have trace proportional to the Weitzenb\"{o}ck
invariant $T%
=S_{\ \mu \nu }^{\rho }T_{\rho }^{\ \mu \nu }$. In order to show
this fact, we can factor out $\sqrt{|g_{\mu \nu }|}$ from
expression (\ref{acciondet}) and use the expansion of the
determinant,
\begin{equation*}
\sqrt{|\mathbb{I}+2\lambda^{-1}\mathbb{F}|}=1+\lambda^{-1}Tr(\mathbb{F})+\mathcal{O}(\lambda^{-2}),
\end{equation*}%
where $\mathbb{F}\equiv \mathcal{F}_{\mu }^{\,\,\nu }$ and
$\mathbb{I}$ is the identity. This last equation expresses the
fact that in any number of dimensions we have
\begin{equation*}
\mathcal{I}_{\mathbf{BIG}}|_{\lambda\rightarrow\infty}=\frac{1}{16\pi
G}\int d^{n}x\;e\;(T-2\Lambda) .
\end{equation*}%
Hence, at the lowest order we retrieve the low energy regime
described by the Einstein theory with cosmological constant
$\Lambda =\lambda (\alpha-1)/[2(A+B)]$.

\section{Non trivial frames for spatially curved FRW Universes}

Due to the fact that general relativity is a theory for the metric,
it is invariant under local Lorentz transformations of the vierbein.
However, the equivalence between Teleparallelism (\ref{lagrangianT})
and GR dynamics, expressed in the Eq.~(\ref{divergence}), implies
that $T$ changes by a boundary term under local Lorentz
transformations. Because of this reason, the teleparallel equivalent
of GR does not provide the manifold with a parallelization but only
with a metric. On the contrary, in a $f(T)$ theory the
\textquotedblleft boundary term" in $T$ will remain encapsulated
inside the function $f$. This means that a $f(T)$ theory is not
invariant under local Lorentz transformations of the vierbein. So, a
$f(T)$ theory will determinate the vierbein field almost completely
(up to global Lorentz transformations). In other words, a $f(T)$
theory will describe more degrees of freedom than the teleparallel
equivalent of GR. This is an important issue in the search for
solutions to the $f(T)$ dynamical equations, since pairs of vierbein
fields connected by local Lorentz transformations (i.e., leading to
the same metric tensor) are \emph{inequivalent} from the point of
view of the theory. Clearly, the same considerations are also valid
for Born-Infeld gravity because the determinantal action
(\ref{acciondet}) is constructed with quadratic combinations of the
torsion tensor, which fails to be invariant under a local Lorentz
change of the tetrad.

Due to the lack of local Lorentz invariance, the vierbein fields
solving the equations of theories with absolute parallelism are
harder to obtain than the metric field. In fact, the considerations
of symmetry about the form of the metric can be of little help to
determine which among a set of locally Lorentz related frames is the
right one to solve the dynamical equations. As an example, we will
find here the adequate frames in order to deal with FRW Universes
with non flat spatial sections. For these geometries, the line
element can be described in hyper-spherical coordinates as
\begin{equation}
ds^{2}=dt^{2}-\kappa^{2}\,a^{2}(t)\,\ [d(\kappa\psi )^{2}+\sin
^{2}(\kappa\psi )\ (d\theta ^{2}+\sin ^{2}\theta \ d\phi ^{2})],
\label{metesferica}
\end{equation}
where $\kappa=1$ for the closed Universe and $\kappa=i$ for the open
Universe. Here, one is tempted to think that the vierbein that
solves the dynamical equations could have the form
\begin{eqnarray}
&&e^{0^{\prime}}=dt,\notag\\
&&e^{1^{\prime}}=\kappa\,a(t)\,d(\kappa\psi ),\notag\\
&&e^{2^{\prime}}=\kappa\,a(t)\,\sin (\kappa\psi )\,d\theta,\notag \\
&&e^{3^{\prime}}=\kappa\,a(t)\,\sin (\kappa\psi )\,\sin \theta
\,d\phi. \label{naive}
\end{eqnarray}
However, this choice turn out to be incompatible with the field
equations coming from the actions (\ref{action}) and
(\ref{acciondet}), because the vierbein (\ref{naive}) does not
correctly parallelize the spacetime. The symptom that the choice
(\ref{naive}) will not work is the form acquired by the
Weitzenb\"{o}ck invariant in such case, which turn out to be
\begin{equation}
T=2\,[(\kappa\,a)^{-2}\,\cot^{2}(\kappa\psi)-3H^2].
\end{equation}
This form of $T$ would be unable of giving a proper reduced
Lagrangian for the dynamics of the scale factor $a(t)$, as a
consequence of its dependence on $\psi $. This $\psi$-dependent
Weitzenb\"{o}ck invariant is not consistent with the isotropy and
homogeneity of the FRW cosmological models.

We shall here sumarise the technique in order to obtain the
parallelized frames. For details, we refer the reader to
Ref.~\cite{Nos5}.

The procedure starts taking the \emph{naive} frame of Eq.
(\ref{naive}) and performing on it a suitable local Lorentz
transformation. This transformation will locally twist the diagonal
frame (\ref{naive}) to convert it in a global well defined parallel
field. For the closed Universe ($\kappa =1$), the local Lorentz
transformation consist in a local rotation of the spatial tetrad
$e^{1^{\prime}},e^{2^{\prime}},e^{3^{\prime}}$, with Euler angles
$\psi ,\ \theta ,\ \phi $; thus both frames are related via the
Euler matrix
\begin{equation}
e^{a}=\mathcal{R}_{a^{\prime }}^{a}\ e^{a^{\prime }},
\label{rotacion}
\end{equation}
where
\begin{equation*}
\mathcal{R}={\footnotesize \left(
\begin{array}{cccc}
1 & 0 & 0 & 0 \\
0 & 1 & 0 & 0 \\
0 & 0 & \cos \phi  & \sin \phi  \\
0 & 0 & -\sin \phi  & \cos \phi
\end{array}%
\right) \left(
\begin{array}{cccc}
1 & 0 & 0 & 0 \\
0 & \cos \theta  & \sin \theta  & 0 \\
0 & -\sin \theta  & \cos \theta  & 0 \\
0 & 0 & 0 & 1%
\end{array}%
\right) \left(
\begin{array}{cccc}
1 & 0 & 0 & 0 \\
0 & 1 & 0 & 0 \\
0 & 0 & \cos \psi  & \sin \psi  \\
0 & 0 & -\sin \psi  & \cos \psi
\end{array}
\right) }.
\end{equation*}
As usual, the angular coordinates range in the intervals $0\leq \phi
< 2\pi $, $0\leq \theta \leq \pi $ and $0\leq \psi \leq \pi $. The
aspect of the so locally rotated frame is now
\begin{eqnarray}
\mathring{E}^{1} &=&- c(\theta)d\psi +\ s(\psi) s( \theta)\ (
c(\psi)
d\theta -\ s(\psi) s(\theta)d\phi )  \notag \\
\mathring{E}^{2} &=& s(\theta) c( \phi)d\psi - s( \psi)[(s(
\psi)s( \phi) - c( \psi) c( \theta)c( \phi) )\ d\theta +(c( \psi)
s( \phi)
 + s( \psi) c( \theta)c( \phi) )\ s(
\theta)\ d\phi ]\notag \\
\mathring{E}^{3} &=&- s( \theta)s( \phi)d\psi - s( \psi) [(s(
\psi) c( \phi) + c( \psi) c( \theta) s( \phi) )\ d\theta +( c(
\psi)
 c( \phi) - s( \psi) c(
\theta) s( \phi) )\ s( \theta) \ d\phi ],\notag \\
\label{autop1}
\end{eqnarray}
where, of course, $s\equiv sin$ and $c\equiv cos$.

A similar procedure is applicable to the open FRW Universe; take the
same rotation (\ref{rotacion}) starting from the spatial part of the
naive vierbein (\ref{naive}) with $\kappa=i$, but now replace the
Euler angle $\psi $ by $i\psi $. The transformed fields read (here
we put $sh\equiv sinh$ and $ch\equiv cosh$)

\begin{eqnarray}
\breve{E}^{1} &=&c(\theta)d\psi +sh(\psi)s(\theta) \ (- ch(\psi)
d\theta +i\ sh( \psi)s(\theta)d\phi )  \notag \\
\breve{E}^{2} &=&-s(\theta)c(\phi)d\psi +sh(\psi)\ [(i\ sh(\psi)
s(\phi) -ch(\psi)c(\theta)c(\phi))\ d\theta +(ch(\psi)s(\phi) +i\
sh(\psi)c(\theta)c(\phi))s(
\theta) \ d\phi ]  \notag \\
\breve{E}^{3} &=&s(\theta)s(\phi)d\psi + sh(\psi) \ [(i\
sh(\psi)c( \phi) + ch(\psi)c(\theta)s(\phi))\ d\theta
+(ch(\psi)c(\phi) -i\ sh(
\psi)c(\theta)s(\phi) )s(\theta) \ d\phi ].\notag \\
\label{autop3}
\end{eqnarray}

Then, as the frame field of 1-forms we take
\begin{equation}
e^{0}=dt;\,\,\,e^{1}=a(t)\ E^{1};\,\,\,e^{2}=a(t)\
E^{2};\,\,\,e^{3}=a(t)\ E^{3}, \label{autop4}
\end{equation}
where the $E^{i}$'s allude to the dreibein $\mathring{E}^{i}$ and
$\breve{E}^{i}$, $i=1,2,3$, of Eqs.~(\ref{autop1}) and
(\ref{autop3}) respectively. The fact that the naive vierbein
(\ref{naive}) is connected with (\ref{autop4}) by a local rotation
guarantees that the latter still describes the spatially curved FRW
metric given in Eq.~(\ref{metesferica}).

The Weitzenb\"{o}ck invariant associated to the vierbein (\ref{autop4}) is%
\begin{equation}\label{invarcerrado}
T=6\,(\pm a^{-2}-H^2),
\end{equation}%
where the positive sign corresponds to the closed case and the
negative to the open one, and $H=\dot{a}/a$ is the Hubble rate
\footnote{Nicely, the flat, closed and open cases ($K=0,1,-1$) can
be summarized in the expression $T=6\,(K\, a^{-2}-H^2)$. See
Ref.~\cite{Nos} for the flat case.}. The absence of angular
vestiges in (\ref{invarcerrado}) foretell that the vierbein
(\ref{autop4}) will be adequate to solve the dynamical equations.
Let us prove this statement first for the $f(T)$ schemes
represented by the action (\ref{action}).

For this purpose, we must replace the vierbein (\ref{autop4}) in the
field equations coming from the variation of the action
(\ref{action}) with respect to the components of the vierbein. When
varying with respect to the temporal component we obtain the
modified version of Friedmann equation, which turns out to be
\begin{equation}
12H^{2}f^{\prime }(T)+f(T)=16\pi G\rho . \label{friedcerrada}
\end{equation}
The non trivial equations for the spatial sector are equal to
\begin{equation}
4(\pm a^{-2}+\dot{H})(12H^{2}f^{\prime \prime }(T)+f^{\prime }(%
T))-f(T)- 4 f^{\prime }(T)\ (2\dot{H}+3H^{2})=16\pi Gp.
\label{espcerradas}
\end{equation}
In the right hand side of the equations appear the energy density
$\rho$ and the pressure $p$ of the fluid that fills the spacetime
in a homogeneous and isotropic way, i.e., the energy-momentum
tensor reads $T_{\mu}^{\nu}=diag(\rho,-p,-p,-p)$ in the comoving
frame. Note that Eq.~(\ref{friedcerrada}) is of first order in
time derivatives of the scale factor, irrespective of the function
$f$. Equations (\ref{friedcerrada}) and (\ref{espcerradas}) are
two differential equations for just one unknown function $a(t)$;
so, they cannot be independent. The way to see that this is indeed
the case, is to take the time derivative of (\ref{friedcerrada})
and combine it with the conservation equation,
\begin{equation}
\dot{\rho}=-3H(\rho+p),  \label{densidad de energia}
\end{equation}
to obtain Eq.~(\ref{espcerradas}). Conversely, if the system
(\ref{friedcerrada}) and (\ref{espcerradas}) is consistent, then the
conservation of energy in the matter sector is given automatically
and Eq.~(\ref{densidad de energia}) holds. A further
characterization of the scale factor dynamics depends on the
particular choice of the function $f(T)$, but the general
considerations about the correctness of the frames (\ref{autop4})
are valid for all $f$'s.

\bigskip

The frames (\ref{autop4}) are also valid for other theories with
absolute parallelism, because they reflect the procedure of
parallelization performed in the FRW manifold. We shall use them now
in order to study the dynamics of the curved FRW models in four
dimensional Born-Infeld gravity. For simplicity, we will work in the
subspace of the parameter space $(A,B)$ defined by the condition
$2A+B=0$ (see Eq.~(\ref{tensorF})).

The dynamics of the scale factor is encoded in the modified
Friedmann equation, which emerges from the replacement of the
ansatz (\ref{autop4}) in the motion equation coming from varying
the action (\ref{acciondet}) with $n=4$ respect the temporal
component of the tetrad $e^{0}$, and it looks
\begin{equation}\label{00k=1}
\frac{\mid 1\pm\frac{1}{\lambda a^2}\mid ^{3/2}}{\sqrt{1-\frac{12
H^2}{\lambda}}}-1=\frac{16 \pi G}{\lambda}\, \rho.
\end{equation}
In the last equation, the signs $\pm$ refer again to the close and
open cases respectively, and the bars mean absolute value. Again,
the equation coming from the spatial sector is just the time
derivative of (\ref{00k=1}) combined with the conservation
equation (\ref{densidad de energia}), so we can ignore it in the
analysis.

An exact solution of the motion equation (\ref{00k=1}) is beyond our
present scope. However, the main physical features can be derived if
one thinks about Eq.~(\ref{00k=1}) as an energy conservation
equation of the type
\begin{equation}
\dot{y}^2+V(y)=0\,\,\,\,\,\,y=\frac{a}{a_{0}}. \label{efpotk=1}
\end{equation}
The potential $V(y)$ can be found easily, and it can be written as
\begin{equation}
V(y)=-\frac{\lambda}{12}y^2\Big[1-\frac{(1\pm
\gamma\,y^{-2})^3}{(1+\beta\,y^{-3(1+\omega)})^2}\Big],
\label{potbik=1}
\end{equation}
where we have defined the two constants $\gamma=1/\lambda a_{0}^2$
and $\beta=16 \pi G \rho_{0}/\lambda$. Additionally, the two
constants $a_{0}$ and $\rho_{0}$ indicate present date values of the
involved quantities, and $\omega$ is the barotropic index of the
perfect fluid with state equation $p=\omega \rho$.

\begin{figure}[h]
\begin{center}
\includegraphics[width=8cm]{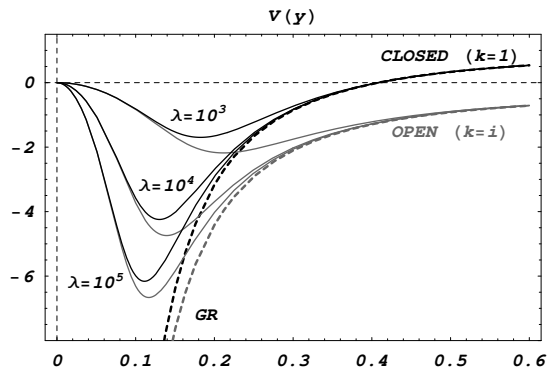}
\caption{\small{Potentials for open (grey) and closed (black)
radiation-dominated FRW Universes in Born-Infeld gravity, as emerge
from Eq.~(\ref{potbik=1}), for several values of the BI parameter
$\lambda$. The dashed curves correspond to the low energy limit of
the theory, i.e., to GR.}} \label{potenciales}
\end{center}
\end{figure}

In Fig. \ref{potenciales} we can see the potentials (\ref{potbik=1})
for a radiation-dominated Universe (i.e., $\omega=1/3$), with
$a_{0}=16 \pi G \rho_{0}=1$, so $\beta=\gamma=\lambda^{-1}$. The
curves in black refers to the closed case and the ones in grey to
the open case. The behavior of GR is depicted in dashed lines for
both cases. While the GR potentials become divergent as
$y\rightarrow 0$, we see that the BI dynamics behave drastically
different in that (high energy) limit. As the scale factor tends to
zero, the BI potentials go to zero with null derivative. In Fig.
\ref{potenciales}, we display the curves for three nominal values of
the BI parameter, $\lambda=10^3$, $\lambda=10^4$ and $\lambda=10^5$.
Note that the energy level in Eq.~(\ref{efpotk=1}) is null, so the
closed Universe possesses a turning point in the low energy regime,
where the potential is null.
The non divergence of the BI potentials in the regime when
$y\rightarrow 0$ denotes a very nice feature of the theory.
Actually, in that limit we have
\begin{equation*}
V(y)=-\frac{\lambda}{12} y^2+\mathcal{O}(y^4),
\end{equation*}
so the Eq.~(\ref{efpotk=1}) has, in the same limit, the solution
\begin{equation*}
a(t)\propto Exp(H_{max}\,t), \,\,\,\,\,
H_{max}=\sqrt{\frac{\lambda}{12}}.
\end{equation*}
Then, in Born-Infeld cosmology we obtain that the early Universe
is characterized by an inflationary stage of geometrical nature.
The exponential behavior of the scale factor is driven by the BI
parameter through the maximum Hubble rate defined according to
$H_{max}=(\lambda/12)^{1/2}$. The existence of this early
inflationary stage does not require the presence of the inflaton,
and it seems to be an intrinsic feature of the BI regularization
procedure. In fact the spacetime so obtained is regular and there
is not Big Bang singularity in it, because the scale factor tends
to zero when the proper time $t\rightarrow -\infty$
\footnote{Though not shown here, all this properties are shared by
the model with flat spatial section.}.

\section{Final remarks}

In this article we have discussed certain conceptual features
related with the lack of local Lorentz invariance in gravitational
theories with absolute parallelism and illustrated its consequences
by means of two examples. In every gravitational theory possessing
absolute parallelism, the spacetime structure is materialized in the
coframe field $\{e^a\}$ which defines an orthonormal basis in the
cotangent space $T_{p}^{\ast}M$ of the manifold $M$ at each point
$p\in M$. As an example we analyzed the dynamics of FRW models in
the so called $f(T)$ theories, which are enjoying a growing
popularity in our days. When $f(T)=T$, i.e., when one considers
general relativity in Weitzenb\"{o}ck spacetime, the basis $\{e^a\}$
at two different points of the manifold are completely uncorrelated,
and it is not possible to define a global smooth field of bases
unambiguously. This is so because the theory is invariant under the
local Lorentz group acting on the coframes $\{e^a\}$. In turn, when
$f(T)\neq T$, local Lorentz rotations and boosts are not symmetries
of the theory anymore. Because of this lack of local Lorentz
symmetry, the theory picks up a preferential global reference frame
constituted by the coframe field $\{e^a\}$ that solves the field
equations. In such case, the bases at two different points become
strongly correlated in order to realize the parallelization of the
manifold. Nevertheless, the appearance of a preferred reference
frame is a property coming from the symmetries of the spacetime, and
it is not ruled by the specific form of the function $f(T)$. For
instance, when one is dealing with FRW cosmological spacetimes, the
field of frames that will lead to consistent field equations will be
given by (\ref{autop4}), whatever the function $f(T)$ be.

Additionally, the fields (\ref{autop4}) are also valid in more
general theories with absolute parallelism which are not related
with the $f(T)$ schemes, for instance, Born-Infeld gravity
(\ref{acciondet}). In this framework, we have seen that the
behavior of the early Universe is characterized by an inflationary
stage of infinite duration and geometrical character. The
exponential form of the scale factor is ruled by the BI parameter
$\lambda$ and it does not rely on the existence of the inflaton.
Finally, and due to the very nature of the Born-Infeld action, the
spacetime results free of the Big Bang singularity.

\section*{Acknowledgments}
F. Fiorini wants to thank the organizers of the 8th Alexander
Friedmann International Seminar for their hospitality at Rio de
Janeiro. This research was supported by CONICET and Universidad de
Buenos Aires.

\end{document}